\documentclass[11pt]{article}
\newcommand{\xxx}[1]{ [#1]}
\newlength{\dummysp}
\newcommand{\beq}{\begin{equation}}
\newcommand{\eeq}{\end{equation}}
\newcommand{\ben}{\begin{enumerate}}
\newcommand{\een}{\end{enumerate}}

\newcommand{\tr}{\mathop{{\hbox{tr} \, }}\nolimits}
\newcommand{\mtxt}[1]{\mathop{\hbox{{\small #1}}}\nolimits}

\newcommand{\half}{{1 \over 2}}

\newcommand{\beqa}{\begin{eqnarray}}
\newcommand{\eeqa}{\end{eqnarray}}
\newcommand{\nnn}{ \nonumber \\ }
\newcommand{\mod}{{\; \mtxt{mod} \; }}
\newcommand{\p}{{\partial}}
\newcommand{\Lcal}{{\cal L}}

\newcommand{\Zbf}{{{\bf Z}}}

\newcommand{\s}{{\sigma}}

\newcommand{\vev}[1]{{\langle #1 \rangle}}

\newcommand{\ordnt}[1]{{{\cal O}(#1)}}

\newcommand{\gappeq}{\mathrel{\rlap {\raise.5ex\hbox{$>$}}
{\lower.5ex\hbox{$\sim$}}}}
\newcommand{\lappeq}{\mathrel{\rlap{\raise.5ex\hbox{$<$}}
{\lower.5ex\hbox{$\sim$}}}}
\newcommand{\myref}[1]{(\ref{#1})}

\newcommand{\ux}{$U(1)_X$}

\newcommand{\bfe}[1]{{\bf #1 \hspace{3pt}}}
\newcommand{\LamH}{\Lambda_H}
\newcommand{\LamX}{\Lambda_X}
\newcommand{\hc}{{\rm h.c.}}
\newcommand{\lra}{\leftrightarrow}
\def\[{\left[}
\def\]{\right]}
\def\({\left(}
\def\){\right)}

\begin{document}

\begin{titlepage} 

\baselineskip=14pt

\hfill    LBNL-49990

\hfill    UCB-PTH-02/12

\hfill    hep-ph/0204017

\hfill    July, 2002

\begin{center}

\vspace{50pt}

{ \bf \Large CP Violation and Moduli Stabilization
in Heterotic Models}\footnote{Based on a talk given
at the CP Violation Conference, University of Michigan,
Ann Arbor, November 4-18, 2001.}

\end{center}

\vspace{5pt}

\begin{center}
{\sl Joel Giedt}\footnote{E-Mail: {\tt JTGiedt@lbl.gov}}

\end{center}

\vspace{5pt}

\begin{center}

{\it Department of Physics, University of California, \\
and Theoretical Physics Group, 50A-5101, \\
Lawrence Berkeley National Laboratory, Berkeley,
CA 94720 USA.}\footnote{This work was supported in part by the
Director, Office of Science, Office of High Energy and Nuclear
Physics, Division of High Energy Physics of the U.S. Department of
Energy under Contract DE-AC03-76SF00098 and in part by the National
Science Foundation under grant PHY-0098840.  Additional
support was provided by the organizers of the CP Violation
Conference and the University of Michigan, Ann Arbor.}

\end{center}

\vspace{5pt}

\begin{center}

{\bf Abstract}

\end{center}

\vspace{5pt}

The role of moduli stabilization in predictions for
CP violation is examined in the
context of four-dimensional effective supergravity
models obtained from the weakly coupled
heterotic string.  We point out that while
stabilization of compactification moduli has
been studied extensively, the determination
of background values for other scalars
by dynamical means has
not been subjected to the same degree of
scrutiny.  These other complex scalars are
important potential sources of CP violation
and we show in a simple model how their
background values (including complex phases)
may be determined from the
minimization of the supergravity scalar
potential, subject to the constraint of
vanishing cosmological constant.

\vfill

\end{titlepage}

\renewcommand{\thepage}{\roman{page}}
\setcounter{page}{2}
\mbox{ }

\vskip 1in

\begin{center}
{\bf Disclaimer}
\end{center}

\vskip .2in

\begin{scriptsize}
\begin{quotation}
This document was prepared as an account of work sponsored by the United
States Government. Neither the United States Government nor any agency
thereof, nor The Regents of the University of California, nor any of their
employees, makes any warranty, express or implied, or assumes any legal
liability or responsibility for the accuracy, completeness, or usefulness
of any information, apparatus, product, or process disclosed, or represents
that its use would not infringe privately owned rights. Reference herein
to any specific commercial products process, or service by its trade name,
trademark, manufacturer, or otherwise, does not necessarily constitute or
imply its endorsement, recommendation, or favoring by the United States
Government or any agency thereof, or The Regents of the University of
California. The views and opinions of authors expressed herein do not
necessarily state or reflect those of the United States Government or any
agency thereof of The Regents of the University of California and shall
not be used for advertising or product endorsement purposes.
\end{quotation}
\end{scriptsize}

\vskip 2in

\begin{center}
\begin{small}
{\it Lawrence Berkeley Laboratory is an equal opportunity employer.}
\end{small}
\end{center}

\newpage
\renewcommand{\thepage}{\arabic{page}}
\setcounter{page}{1}
\def\thefootnote{\arabic{footnote}}
\setcounter{footnote}{0}

\baselineskip=14pt

It has been argued by Dine et al.~and Choi et al.~\cite{CPGS}
that CP is a gauge
symmetry in string theory and that {\it explicit} breaking
is therefore forbidden both perturbatively and
nonperturbatively.  They have shown that this is
certainly true for weakly coupled heterotic orbifolds,
the principal topic of this note.
Thus, we envision {\it spontaneous}
CP violation through complex scalar
vacuum expectation values ({\it vevs}).

\bfe{Compactification Moduli.}  One possible source is
scalar fields corresponding to {\it compactification
moduli} of the six-dimensional compact space.  These are {\it K\"ahler
moduli} (denoted with a ``T'') and
{\it complex structure moduli} (denoted with a ``U'').
Ib\'a\~nez and L\"ust \cite{IL92} have enumerated the
possibilities for $Z_N$ and $Z_M \times Z_N$ orbifolds,
based on the results of \cite{Z3D}:
\beqa
\mtxt{case 1:} \qquad \Gamma & = & SL(2,Z)^3_T \times SL(2,Z)^n_U \nnn
K &=& -\sum_{I=1}^3 \ln (T^I + \bar T^I)
-\sum_{I=1}^n \ln (U^I + \bar U^I) \nnn
n&=&0: \qquad Z_7, Z_8', \ldots \nnn
n&=&1: \qquad Z_6, Z_8, \ldots \nnn
n&=&3: \qquad Z_2 \times Z_2 \label{ae1} \\
\mtxt{case 2:} \qquad \Gamma & = & SL(2,Z)_T \times SL(2,2,Z)_T \times SL(2,Z)^n_U \nnn
K &=& - \ln (T^1 + \bar T^1) - \ln \det (T + \bar T)
- \sum_{I=1}^n \ln (U^I + \bar U^I) \nnn
& & T \quad {\rm is} \quad 2 \times 2, \qquad
n=0: \quad Z_6', \qquad n=1: \quad Z_4 \label{ae2} \\
\mtxt{case 3:} \qquad \Gamma & = & SL(3,3,Z)_T : \quad Z_3 \nnn
K &=&  - \ln \det (T + \bar T), \quad T \quad {\rm is} \quad 3 \times 3
\label{ae3}
\eeqa
$\Gamma$ is the {\it target-space modular duality group.}
It can be seen that each case has 3 {\it diagonal}
K\"ahler moduli $(I=1,2,3)$:
\beq
T^I \equiv T^{II} 
\label{ae4}
\eeq
which transform under an $SL(2,Z)^3$ subgroup of the
duality group $\Gamma$:
\beqa
T^I &\to& {T'}^I = {a^I T^I - i b^I
\over i c^I T^I + d^I}, \nnn
& & a^I d^I - b^I c^I = 1, \qquad
a^I, b^I, c^I, d^I \in \Zbf
\label{ae5}
\eeqa
Enforcing this symmetry on the field theory limit
leads to {\it modular invariant supergravity.}
This symmetry constrains the stabilization of
these moduli, and hence possible CP violating
phases originating from $\arg(T^I)$ \cite{BKL,Gie01a,Den00,Leb01}.

\bfe{FI induced vevs.}  Cancelation of the trace anomaly
associated with an anomalous
\ux\ by the GS mechanism \cite{GS84}
leads to an FI term \cite{UXR} for
the D-term of \ux:
\beq
D_X = \sum_i K_i q_i^X \phi^i + \xi, \qquad
\xi = {g_H^2 \tr Q_X \over 192 \pi^2} m_P^2
\label{ae6}
\eeq
where $g_H$ is the unified coupling at the
string scale $\LamH$ and $m_P = 1/\sqrt{8\pi G}
= 2.44 \times 10^{18}$ GeV is the reduced Planck
mass.  The fields $\phi^i$ which acquire nonvanishing
vevs will be (following \cite{Gie01a}) referred
to as {\it Xiggs} fields in what follows.

In \cite{Gie02} it was shown that
the presence of a \ux\ factor in the gauge group $G$
is generic for semi-realistic
orbifold models.  For the class
of standard-like orbifolds studied
there, only 7 of 175 models did not have
a \ux.  In the semi-realistic {\it free fermionic}
models \cite{CFN99} a \ux\ is also generic.

In \cite{GG00}, the scalar potential $V$
for SUGRA with a \ux\ was studied for vacuum configuations
satyisfying $\vev{V} = \vev{\p V / \p \phi^i } = 0$.
Supersymmetry breaking was characterized by
\beq
\vev{|W|^2} = |\delta|^2, \quad
\vev{K_{i \bar j} F^i \bar F^{\bar j} }
= \alpha e^{\vev{K}} |\delta|^2
\label{ae7}
\eeq
According to expectations, it was found that 
$\vev{V} = \vev{{\p V / \p \phi^i }} = 0$ together
with reasonable supersymmetry breaking scale
requires
\beq
\vev{D_X} \sim |\delta|^2 \ll |\xi|
\label{ae8}
\eeq
For canonical $K= \sum_i |\phi^i|^2$ and
$\vev{\phi^i} = v^i,$
\beqa
\vev{D_X} &=& \sum_i q_i^X |v_i|^2 + \xi \label{ae9} \\
q_i^X &\sim& 1 \quad \Rightarrow \quad |v^i| \sim \sqrt{|\xi|}
\label{ae10a}
\eeqa

Research in progress \cite{GG02} has shown that
(\ref{ae8},\ref{ae10a}) hold in cases more complicated
than those studied in \cite{GG00}.

Based on \myref{ae10a}, in \cite{Gie02} the \ux\
gauge symmetry breaking scale $\LamX$
was defined as
\beq
\LamX = \sqrt{|\xi|}
\label{ae10b}
\eeq
For the class of models studied there it was found
that for the 168 of 175 cases where $\xi \not= 0$,
\beq
{g_H \over 8.00} \leq {\LamX \over m_P}
\leq {g_H \over 4.63} = {\LamH \over m_P}
\label{ae11}
\eeq
where $\LamH \approx 0.216 \times g_H m_P$ is the
approximate string scale obtained in \cite{Kap88}.
With $g_H \sim 1$ we have that $\LamX \sim 0.1 \times m_P$
is a generic prediction.

The result of this is that nonrenormalizable operators
should contribute {\it significantly} to the
(effective) Yukawa couplings of the lighter
quarks, since they are only down by
$(\LamX / m_P)^n \sim 10^{-n}, n>0$.  Given
$\lambda_{u,d}/\lambda_t \sim 10^{-5}$ after
running to the high scale, it is difficult to believe that
nonrenomalizable operators would
not play a role, generically
speaking.  Operators with $1 \leq n \leq 4$
would typically be present.
Models with flat directions
where this is not the case may be able to
be found; for example, the 7 of 175 without
a \ux\ found in \cite{Gie02}.  However, if
this is not the case generically, one can
take the point of view that these models do not
well represent ``predictions'' of string theory.
That is, one can argue that in the absence of a dynamical
vacuum selection mechanism
in string theory, one should fall back on what
is generic
in extracting ``predictions.''

For these reasons, it was argued in \cite{Gie01a}
that FI induced vevs are
more likely to be the dominant source of CP
violation in string-derived models.  These
vevs are generically complex.
Indeed, \myref{ae9} is completely
``phase-blind,'' leading to
massless pseudoscalars termed {\it D-moduli}
in \cite{GG00}.  Efforts
are underway \cite{GG02} to stabilize
these moduli by including
various terms in $V$ (intentionally) neglected in
\cite{GG00}.  Until this can be achieved, there
is no {\it dynamical} reason to take the
phases to be real.  In \cite{GG00} it was
demonstrated that the generic case leads
to nonzero KM phase.

\bfe{Relevant orders.}
Renormalizing to $\LamX \sim \LamH \sim 10^{17}$ GeV,
\beq
{m_u \over m_t} \sim 10^{-5}
\label{be1}
\eeq
for moderate values of $\tan \beta$.  But for $T^I$ stabilized
at self-dual values, we do not get such a hierarchy
from trilinear Yukawas.  A natural source of such
a hierarchy is the scale $\LamX$.  I.e., nonrenomalizable
superpotential couplings can
give effective Yukawa couplings from Xiggs vevs.
The hierarchy is generically
\beq
{\lambda_{eff} \over \lambda_{tri} } \sim
\( {\LamX \over m_P} \)^n
\eeq
where $n$ counts dimensionality beyond trilinear.  To
get \myref{be1} we need $n \approx 5$.  Thus, the
structure of the Yukawas
requires that we consider operators of rather large
dimension in order to develop {\it predictions}
in a realistic theory.  The smallness
of $m_u, m_d, m_s$ requires that
we work well beyond leading order to
make firm conclusions about,
say, $V_{ub}$ vs. experiment.
Not only should we work to
higher orders in the superpotential
couplings, but also in the
K\"ahler potential since Xiggs
vevs could give kinetic mixing which is not that small.
E.g., $K \ni k_{ijk}\vev{\phi_i}u_j^\dagger u_k,
j \not= k$.
Such higher order terms are not well-understood and are not
protected by nonrenormalization theorems.  For the latter reason
we will generally need to
study supergravity loop corrections to any K\"ahler
coupling extracted from string amplitudes
at the string scale.

Note that all of these effects are highly dependent on
the choice of Xiggs vevs.  Which flat direction do we
lead our expansion about?  The results of \cite{Gie01a}
seem to suggest that the degeneracy is so great that
we likely have the flexibility to
tune the KM phase to any value we like.  String theory
has provided us with an empirical framework
to match experimental data, while at the same time
rendering our description quantum
mechanically consistent in
the ultraviolet limit with gravity included.
But can we do better? Can we extract predictions?

Consider what Bin\'etruy, Gaillard and Wu (BGW)
did to extract predictions for
the T-moduli phase \cite{BGW}.  They included nonperturbative
effects from the hidden sector, coupled $T^I$ to
the gaugino condensate superfield $U$ in
the Veneziano-Yankielowicz lagrangian $\Lcal_{VY}$
and imposed reparameterization symmetries.  We must
envision a similar analysis to stabilize the
D-moduli if we are to stabilize them and predict
Xiggs phases.  The mixed anomaly
\beq
\tr Q_X \not= 0 \quad \Rightarrow \quad
\tr T^a T^a Q_X \not= 0
\eeq
implies Xiggs couplings through gauge interactions
with the hidden sector.  A study of these effects
is in progress.

\bfe{Phase predictions from D-moduli stabilization.}
This is a modification of the linear multiplet
toy model of \cite{GG00}.  The desire is to lift
vacuum degeneracy by coupling D-moduli to matter condensates
of the hidden sector condensing group $G_C$.  As just mentioned,
such
couplings are expected from the mixed trace anomaly
matching condition $\tr T^a T^a Q_X \not= 0,$
where $T^a$ is a generator of $G_C$.  The notation
in what follows is defined in \cite{GG00}.

We continue to have
\beqa
K &=& k(L) + G(A,B,\Phi,\bar A,\bar B,\bar \Phi) \nnn
k(L) &=& \ln L + g(L) \nnn
G &=& \sum_i |A_i|^2 + \sum_i |B_i|^2 + \sum_i |\Phi_i|^2
\label{ce1}
\eeqa
On the other hand the superpotential now takes the
form
\beqa
W(A,B,\Phi,\Pi) &=& \hat W(A,B,\Phi) + \breve W(\Phi,\Pi) \nnn
\hat W(A,B,\Phi) &=& \lambda_{ijk} A_i B_j \Phi_k \nnn
\breve W(\Phi,\Pi) &=& c_\alpha(\Phi) \Pi_\alpha
\label{ce2}
\eeqa
The functional $c_\alpha(\Phi)$ is left
unspecified at this point.  The fields $\Pi_\alpha$
are chiral superfields corresponding to hidden
sector matter condensate operators.

We implement dynamical supersymmetry breaking
through the form of the Vene\-zi\-ano-Yan\-ki\-elo\-wicz lagrangian
assumed in \cite{BGW}
\beq
\Lcal_{VY} = \int {E \over 8R} U \[ b' \ln (e^{K/2} U)
+ \sum_\alpha b^\alpha \ln \Pi^\alpha \] + \hc
\label{ce3}
\eeq
where $U$ is the chiral superfield corresponding to the
hidden sector gaugino bilinear condensate operator.
We have no T-moduli appearing explicitly, no threshold
corrections, and the only GS term is the one required 
to cancel the \ux\ anomaly.
Following the BGW formulation with these simplifications
one obtains for the scalar potential
\beqa
V &=& \half \( {2 \ell \over 1+f} \) \sum_a D_a D_a
+ (\ell g' - 2) \left| {b' u \over 4} - e^{K/2} W \right|^2 \nnn
&& + \left|e^{K/2} (\hat W_I + W G_I) - {b' u \over 4} G_I \right|^2 \nnn
&& + \( {1+ \ell g' \over 16 \ell^2} \)
\[ (1+2 \ell b') |u|^2 - \ell e^{K/2} (W \bar u + \bar W u) \]
\label{ce4}
\eeqa
Here, $\ell = L|, u=U|$.

We restrict our attention to the
submanifold of the vacuum manifold where $D_X$ is the
only nonvanishing D-term, and $\vev{a_i}=\vev{b_i}=0$,
with $a_i=A_i|$, $b_i=B_i$.  In this case the vacuum
image of $V$ is, after straightforward manipulations,
given by
\beqa
V&=& \half g_H^2 D_X^2 + \hat V \label{ce5} \\
\hat V & = & e^K \sigma^2
   \[ b_c^2 (v^2-2+\ell g')
   + \( { 1+\ell g' \over 2 \ell^2} \)
   (2+3 \ell b' + \ell b_c) \]
\label{ce6} 
\eeqa
where we take $\ell$ at its vev and
\beqa
v_i &=& \vev{\phi_i}, \qquad v=\[ \sum |v_i|^2 \]^{1/2} \nnn
D_X &=& \sum_i q_i |v_i|^2 + \xi, \qquad
g_H^2 = {2 \ell \over 1 + f} \nnn
K &=& k(\ell) + v^2, \qquad
b_c = b' + \sum_\alpha b^\alpha \nnn
\s & = & {1 \over 4}
   \exp \[ - {1 \over b_c g_H^2} -  {b' \over b_c} \] 
   \prod_\alpha \left| {4 c_\alpha(v)
   \over  b^\alpha } \right|^{ b^\alpha / b_c}
\label{ce7} 
\eeqa

From these expressions it is not hard to work out
$V_i = \p V/\p v_i$:
\beq
V_i  = \bar v_i \[ g_H^2 D_X q_i + \s^2 b_c^2 e^K + \hat V \]
+ \mu_i \hat V
\label{ce8}
\eeq
where
\beq
\mu_i = \sum_\alpha {b^\alpha \over b_c} {\p \over \p v_i}
\ln c_\alpha (v)
\label{ce9} 
\eeq
Notice that all of the quantities in \myref{ce8} are real except
$\bar v_i$ and $\mu_i$.  Thus, the phase of $v_i$ will
be related to $\mu_i$.  We expect that
this provides the necessary constraint
to lift pseudoscalar D-moduli. More precisely,
\beq
\arg v_i = - \arg \mu_i \mod \pi
\label{ce10} 
\eeq

For minimization with vanishing cosmological constant
we require $V= V_i = 0$, which implies $\hat V \sim -D_X^2$.
To have low scale supersymmetry breaking we demand $|D_X|
\ll 1$, so from \myref{ce8} $|D_X| \sim \s^2 \ll 1$.
Then for $v_i \not=0$ eq. \myref{ce8} implies
\beq
g_H^2 D_X q_i + b_C^2 \s^2 e^K = \ordnt{\s^4}
\label{ce11} 
\eeq
unless $v_i \lappeq \s^2$.  Thus the delicate cancelation
of $\ordnt{\s^2}$ terms implied by \myref{ce10} can, for reasonable
splittings of $q_i$, only be achieved for {\it one choice}
of charge $q_i$, just as was the case in \cite{GG00}.
Since the term $\mu_i \hat V$ in \myref{ce8} is
roughly $\ordnt{\s^4}$, the result of \cite{GG00} for
{\it which} $q_i$ gets vevs $v_i \gg \s^2$ is unchanged:
the minimum $q_i$ must be the set getting $v_i \gg \s^2$
vevs for positive mass-squared.  To conclude,
\beqa
|v_i| &\sim &
\left\{
\begin{array}{l}
\LamX \qquad q_i = -q \\
\s \qquad q_i \not= -q
\end{array}
\right. \nnn
q &=& - \min \{ q_i \}, \quad \LamX = \sqrt{\xi}
\label{ce12} 
\eeqa
since $D_X = \sum_i q_i |v_i|^2 + \xi \sim \s^2 \ll \xi$.
From \myref{ce8} we have
\beq
v^2 = {\xi \over q} - {b_C^2 \s^2 e^K \over q^2 g_H^2 }
+ \ordnt{\s^4}
\label{ce13} 
\eeq

We next suppose in \myref{ce2}
\beq
c_\alpha(v) = \sum_A c_{\alpha A}(v), \qquad
c_{\alpha A}(v) = \lambda_{\alpha A} \prod_i (v_i)^{p_{iA}^\alpha} .
\label{ce14}
\eeq
Then it is easy to check that \myref{ce9}
yields
\beq
v_i \mu_i = \sum_\alpha {b^\alpha \over b_c}
{ \sum_A p_{iA}^\alpha c_{\alpha A}(v) \over
\sum_A c_{\alpha A}(v) }
\label{ce15}
\eeq
Consequently we can rewrite the minimization constraint
which follows from \myref{ce8} as (exactly)
\beq
0 = |v_i|^2 \[ g_H^2 D_X q_i + b_c^2 \s^2 e^K + \hat V \]
+ \hat V \sum_\alpha {b^\alpha \over b_c}
{ \sum_A p_{iA}^\alpha c_{\alpha A}(v) \over
\sum_A c_{\alpha A}(v) }
\label{ce16}
\eeq
In the case where the sum in \myref{ce14} has only a single
term, the $c_{\alpha A}(v)$ cancel in \myref{ce16} and
no phase constraints exist.  Thus, a non-monomial polynomial assumption
for $c_\alpha(v)$ is required for phase stabilization.

As an example, consider the case of only two fields
$\phi^1,\phi^2$ of charges $q_1=q_2 \equiv -q$ and a single
matter condensate field with superpotential coupling
\beq
\breve W(\phi,\pi) = c(\phi) \pi, \qquad
c(\phi) = \lambda_1 \phi_1 + \lambda_2 \phi_2
\label{ce17}
\eeq
We can go to U-gauge by writing
\beq
\pmatrix{\phi_1(x) \cr \phi_2(x)} =
e^{i (\theta(x) + \varphi_+)}
\pmatrix{ e^{i \varphi_-} \cos \eta & e^{i \varphi_-} \sin \eta \cr
-e^{-i \varphi_-} \sin \eta & e^{-i \varphi_-} \cos \eta }
\pmatrix{d(x) \cr h(x) + v}
\label{ce18}
\eeq
where $\vev{\theta(x)}=\vev{d(x)}=\vev{h(x)}=0$ and
\beqa
v_1 &=& e^{i\varphi_1} v \cos \eta, \qquad
v_2 = e^{i\varphi_2} v \sin \eta, \nnn
\varphi_\pm &=& \half (\varphi_1 \pm \varphi_2)
\label{ce19}
\eeqa
Here, $\theta(x) + \varphi_+$ is eaten by the \ux\
vector boson, while the scalar $h(x)+v$ acquires a $\LamX$
mass and fills out the massive \ux\ vector multiplet.
$\varphi_-$ is the phase we would like to stabilize and
$d(x)$ is the complex scalar D-modulus field.  $\eta$ is the
mixing angle to the mass eigenstate basis which
we would also like to stabilize.  It is not hard to
check that \myref{ce8} gives
\beq
0 = b_c ( \lambda_1 |v_1|^2 + \lambda_2 \bar v_1 v_2)
(- q g_H^2 D_X + b_c^2 \s^2 e^K + \hat V) + b^\alpha \lambda_1 \hat V
\label{ce20}
\eeq
and a similar equation with $1 \lra 2$, and then 2 conjugate
equations.  Manipulations on these four equations lead
simply to
\beq
{v_1 \bar v_2 \over \bar v_1 v_2} = 
{\bar \lambda_1 \lambda_2 \over \lambda_1 \bar \lambda_2}
\Rightarrow
\varphi_1-\varphi_2 = 2 \varphi_- =
\arg({\lambda_2 \over \lambda_1}) \mod \pi
\label{ce21}
\eeq
It is also straightforward to check
\beq
\sin^2 \eta = {b^\alpha \hat V (|\lambda_1|^2 - |\lambda_2|^2)
+ b_c v^2 |\lambda_1|^2 (-q g_H^2 D_X + b_c \s^2 e^K + \hat V)
\over 2 b_c v^2 |\lambda_2|^2
(-q g_H^2 D_X + b_c \s^2 e^K + \hat V)}
\label{ce22}
\eeq
Thus $d(x)$ is stabilized and the phase and mixing
are determined.  In particular, the phase $\varphi_-$
is not an independent source of CP violation, but
is determined by whatever mechanism determines the
phases of $\lambda_i$.

If we embed this toy into a string-inspired model,
the source of $\arg(\lambda_2/\lambda_1)$ would
be the phase of T-moduli.  For example, suppose the
modular weight of $\pi$ is $q_\pi^I$
while for the fields $\phi_{1,2}$ we have $q_{1,2}^I$.
Then treating $\phi_{1,2},\pi$ as untwisted fields
we have
\beq
\lambda_i \sim \prod_I [\eta(t^I)]^{2(q_i^I + q_\pi^I -1)}
\qquad (i=1,2)
\label{ce23}
\eeq
Then
\beq
{\lambda_2 \over \lambda_1} \sim
\lambda_i \sim \prod_I [\eta(t^I)]^{2(q_2^I - q_1^I)}
\label{ce24}
\eeq
From \myref{ce21} we have
\beq
\varphi_1 - \varphi_2 = 2 \sum_I (q_2^I - q_1^I)
\arg(\eta(t^I)) \mod \pi
\label{ce25}
\eeq
If $\phi_{1,2}$ are untwisted with different
compact space $SO(6)$ weights ({\it H-momenta}),
say $(1,0,0)$ and $(0,1,0)$ resp., then
\beq
\varphi_1 - \varphi_2 = 2 \[ \arg(\eta(t^2))
- \arg(\eta(t^1)) \] \mod \pi
\label{ce26}
\eeq
For $t^1 = e^{i\pi /6}$, $t^2=1$,
\beq
\arg(\eta(t^1)) = - \pi/24, \qquad
\arg(\eta(t^2)) = 0
\label{ce27}
\eeq
and hence
\beq
\varphi_1 - \varphi_2 = {\pi \over 12} \mod \pi.
\label{ce28}
\eeq
In the case of twisted fields $\phi_1$ or $\phi_2$,
the linear coupling in \myref{ce17} implies that
$\pi$ is also twisted by the point group selection
rule.  In this case a major revision to \myref{ce3}
would be required because of mixing under $SL(2,Z)_T^3$.
A mixing of $\pi$ with other operators would not
give the right sort of 
anomalous modular transformation for $\Lcal_{YV}$
which could be canceled by $\Lcal_{GS}$.

\vspace{20pt}

\noindent {\bf \Large Acknowledgements}
The author would like to thank the organizers of the
CP Violation Conference at the University of Michigan,
Ann Arbor.  Their hospitality is very much appreciated
and was of assistance in the preparation of this work.
I would further like to express my appreciation for
my fruitful collaboration with Mary K.~Gaillard, and
would like to acknowledge her role in the development
of ideas reported here.  Conversations with
Brent D.~Nelson, Thomas Dent and Gordon Kane
also proved influential on the ideas expressed
here.

This work was supported in part by the
Director, Office of Science, Office of High Energy and Nuclear
Physics, Division of High Energy Physics of the U.S. Department of
Energy under Contract DE-AC03-76SF00098 and in part by the National
Science Foundation under grant PHY-0098840.  Additional
support was provided by the organizers of the CP Violation
Conference and the University of Michigan, Ann Arbor.

\end{document}